\documentclass{article}
\usepackage{arxiv}

\AtBeginDocument{%
  }

\usepackage[utf8]{inputenc} 
\usepackage[T1]{fontenc}    
\usepackage{hyperref}       
\usepackage{url}            
\usepackage{booktabs}       
\usepackage{amsfonts}       
\usepackage{nicefrac}       
\usepackage{microtype}      
\usepackage{lipsum}
\usepackage{graphicx}
\usepackage{subcaption}  
\usepackage{amsmath}
\usepackage{amssymb}
\usepackage{wrapfig}

\usepackage{longtable,booktabs}

\begin{document}
\setlength{\textfloatsep}{10pt plus 1pt minus 5pt} 
\widowpenalty=10000 
\title{Uncovering the Limitations of Query Performance Prediction: Failures, Insights, and Implications for Selective Query Processing}
\title{Reevaluating Query Performance Prediction: Challenges in Sparse and Dense Retrieval Systems}
\title{The Limitations of Query Performance Prediction: Insights and Challenges Across Retrieval Paradigms}
\title{The Fragility of Query Performance Prediction Across Rankers and Collections}
\title{Uncovering the Limitations of Query Performance Prediction: Failures, Insights, and Implications for Selective Query Processing - 
}

\author{
  Adrian-Gabriel Chifu \\
  Aix Marseille Université, Université de Toulon, CNRS, LIS \\
  Marseille, France \\
  \texttt{adrian.chifu@univ-amu.fr} \\
  \And
  Sébastien Déjean \\
  IMT, UMR5219 CNRS, UPS, Univ. de Toulouse \\
  Toulouse, France \\
  \texttt{sebastien.dejean@math.univ-toulouse.fr} \\
  \And
  Josiane Mothe \\
  IRIT, UMR5505 CNRS, Université de Toulouse, INSPE, UT2J \\
  Toulouse, France \\
  \texttt{josiane.mothe@irit.fr} \\
  \And
  Moncef Garouani \\
  IRIT, UMR5505 CNRS, Université Toulouse Capitole, UT1 \\
  Toulouse, France \\
  \texttt{moncef.garouani@irit.fr} \\
  \And
  Diego Ortiz \\
  IRIT, UMR5505 CNRS \\
  Toulouse, France \\
  \texttt{diego.ortiz@irit.fr} \\
  \And
  Md Zia Ullah \\
  Edinburgh Napier University \\
  Edinburgh, UK \\
  \texttt{m.ullah@napier.ac.uk} \\
}


\maketitle
\begin{abstract}
 Query Performance Prediction (QPP) estimates retrieval systems effectiveness for a given query, offering valuable insights for search effectiveness and query processing. Despite extensive research,  QPPs face critical challenges in generalizing  across diverse retrieval paradigms and collections. This paper provides a comprehensive evaluation  of state-of-the-art QPPs (e.g., NQC, UQC), LETOR-based features, and newly explored dense-based predictors. Using diverse  sparse rankers (BM25, DFree without and with query expansion) and hybrid or dense (SPLADE and ColBert) rankers and diverse test collections—ROBUST, GOV2, WT10G, and MS MARCO—we investigate the relationships between predicted and actual performance, with a focus on generalization and robustness. Results show significant variability in  predictors accuracy, with collections as the main factor and rankers next.  Some sparse predictors perform somehow on some collections (TREC ROBUST and GOV2) but do not generalise to other collections (WT10G and MS-MARCO).   While some predictors show promise in specific scenarios, their overall limitations constrain their utility for applications. We show that QPP-driven selective query processing offers only marginal gains, emphasizing the need for improved predictors that generalize across collections, align with dense retrieval architectures and are useful for downstream applications. 
 We will publicly release our data and code following acceptance\footnote{\url{https://anonymous.4open.science/r/UncoveringTheLimitationsofQPP-346C/README.md}}.

 \begin{figure}
  \includegraphics[width=0.24\textwidth]{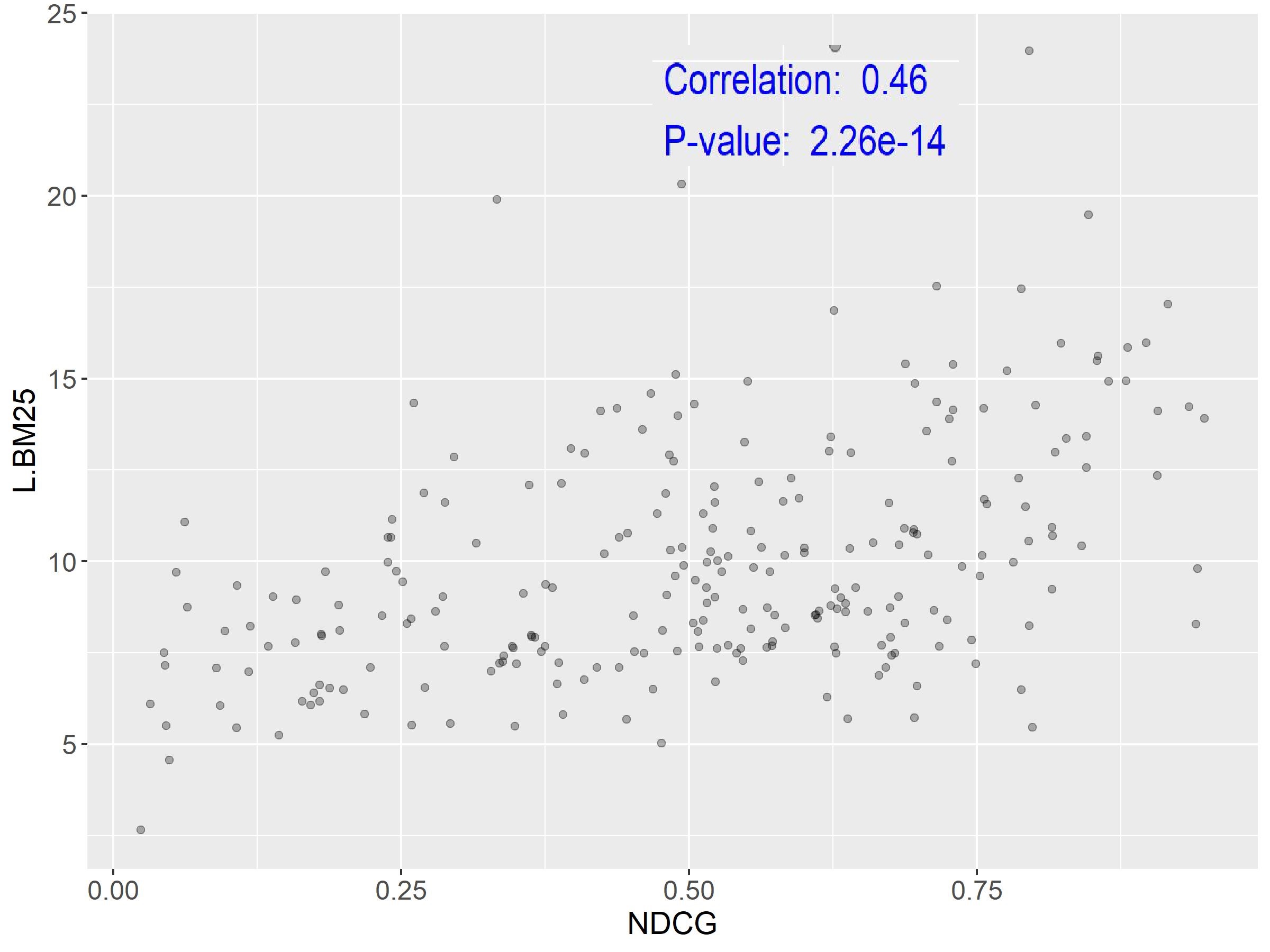}
    \hspace{-0.5mm}%
\includegraphics[width=0.24\textwidth]{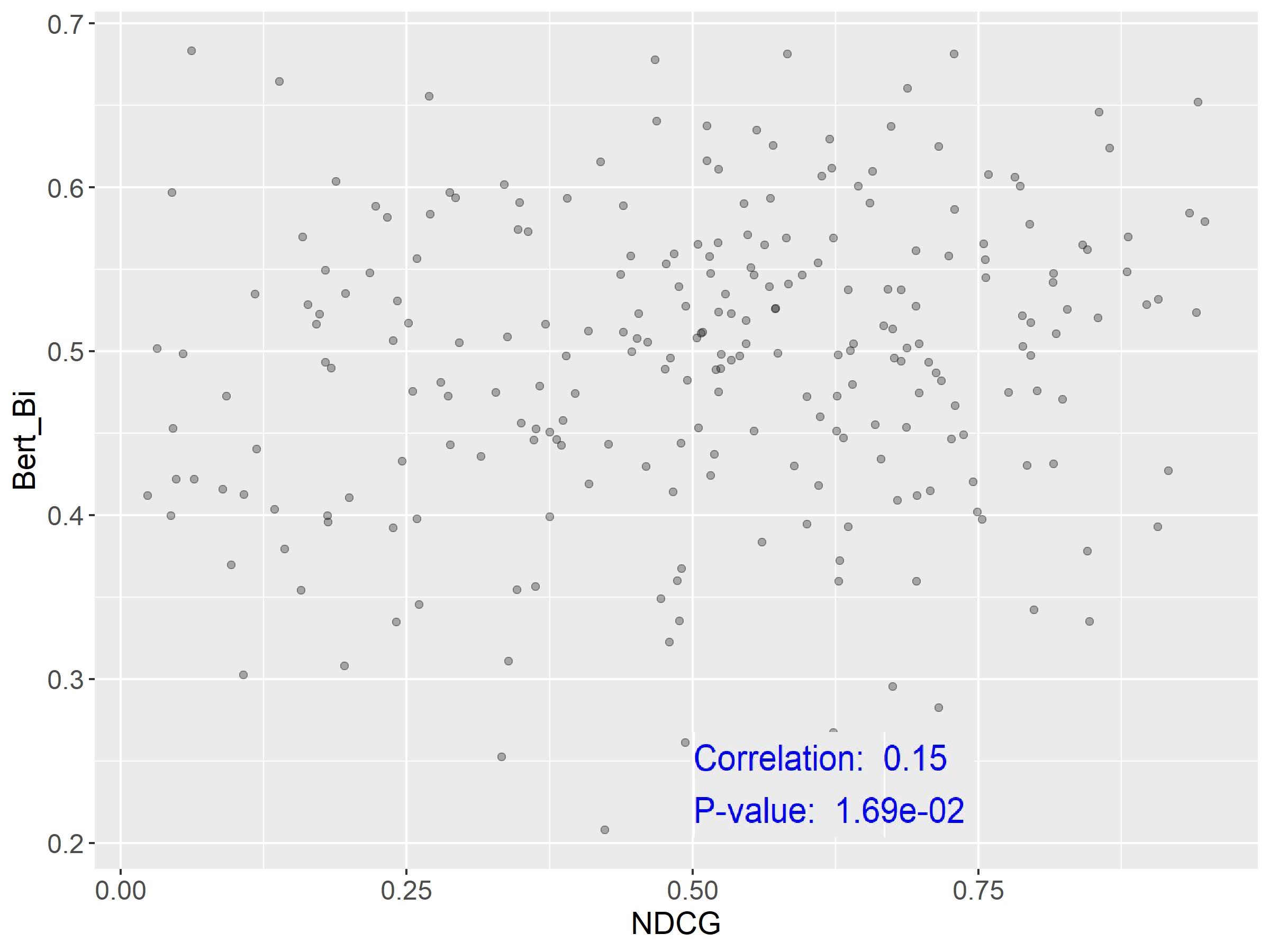}
  \hspace{-0.5mm}%
  \includegraphics[width=0.24\textwidth]{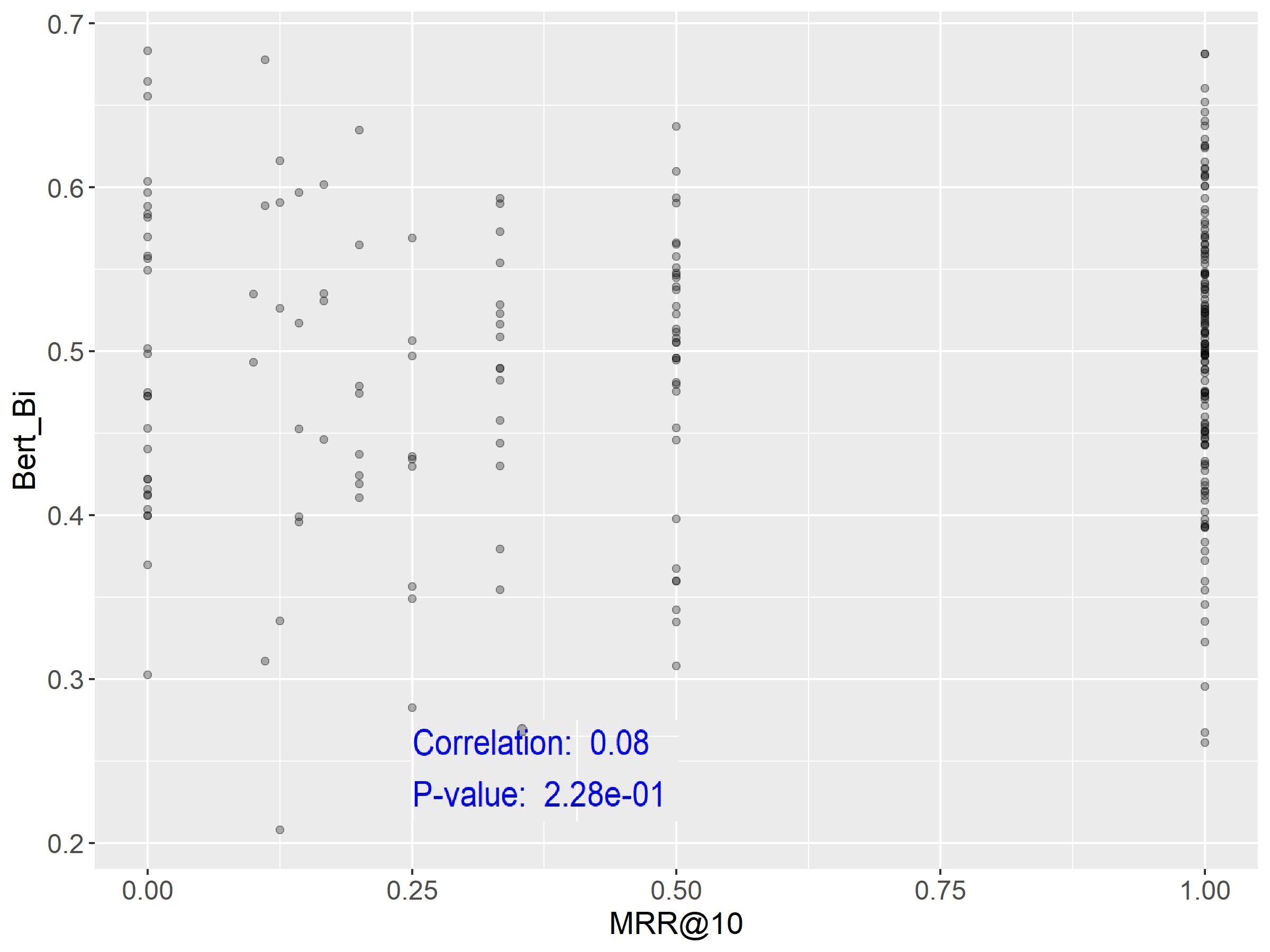}
  \hspace{-0.5mm}%
    \includegraphics[width=0.24\textwidth]{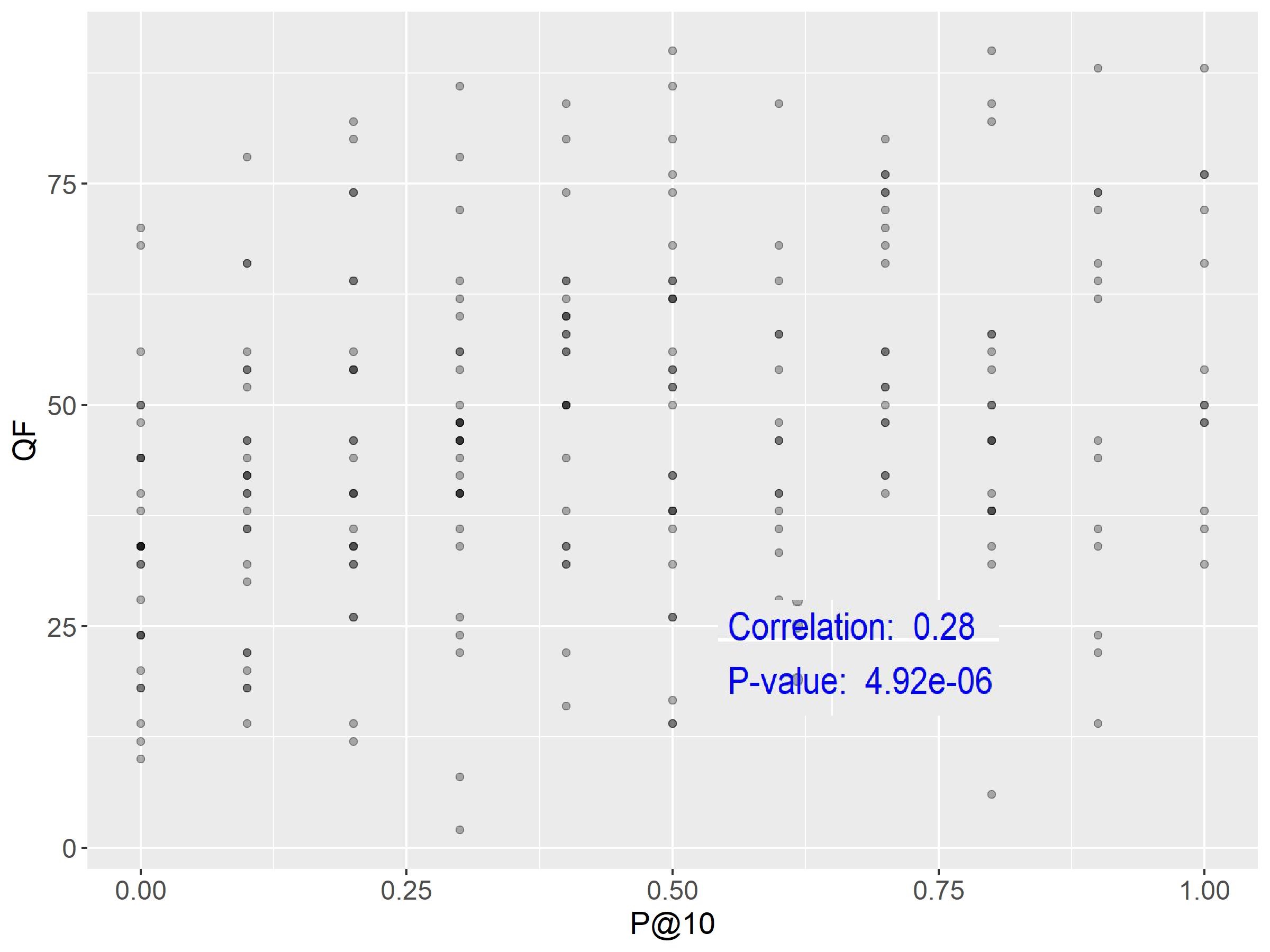}
\caption{How well do existing QPP methods perform across diverse retrieval paradigms and collections? Relationships between predicted values and actual values, as well as Pearson correlations and p-values exhibit significant variability, highlighting the challenges of generalization and robustness across settings.}
  \label{fig:teaser}
\end{figure}

\end{abstract}


\section{Introduction and Motivation}
Query Performance Prediction (QPP)  in information retrieval (IR)  aims at estimating the effectiveness of retrieval systems for a given query before results are retrieved~\cite{carmel2010estimating,dejean2020forward,faggioli2023query,arabzadeh2024query}. Despite decades of research, QPP remains an open problem, as existing methods struggle to generalize across diverse retrieval paradigms and collections. Most studies focus on developing predictors that correlate with actual system performance or minimize prediction error. However, fewer studies address downstream applications, such as selective query expansion~\cite{amati2004query,datta2024deep}, selective query processing~\cite{deveaud2018learningLong} or query-specific variable depth pooling~\cite{ganguly2023query}, which could enhance retrieval efficiency and relevance. The emergence of dense rankers that leverage contextual embeddings, such as SPLADE and ColBERT, introduces new challenges for traditional QPP approaches.

Sparse rankers like BM25 rely on exact term matching and term frequency statistics, which is robust across collections.. In contrast, dense rankers use semantic embeddings to capture richer contextual relationships. These contrasting paradigms introduce complexities for QPP, as state-of-the-art (SOTA) predictors (e.g., NQC \cite{shtok2009predicting}) and LETOR-based features were primarily designed for sparse systems and may fail to capture the nuances of dense retrieval. On the other hand, some predictors have been defined for dense contexts (e.g. BERT-based) but not evaluated on sparse contexts.

This paper provides a comprehensive analysis of QPP methods, including SOTA, LETOR-based and dense-based predictors, evaluated across four datasets: TREC ROBUST, GOV2, WT10G, and MS MARCO on different performance measures (NDCG, MAP, and P@10). Using these datasets, we analyse QPP performance for sparse rankers (BM25, DFree~\cite{amati2002probabilistic}, and their query expansion variants) and dense rankers (SPLADE, ColBERT). Through this comprehensive analysis, we aim to uncover the factors that most significantly impact QPP accuracy and compare the QPP predictors in different settings. This is a pioneer study than encompasses that number of factors and modalities; also reporting failures. 

Our results reveal several key findings:

\begin{itemize}
    \item Predictors are not highly correlated with performance measures, SOTA sparse predictors are even weaker in dense retrieval contexts;
    \item Incorporating dense-based predictors does not solve the problem for dense retrieval and do not generalize in sparse contexts;
    \item Predictors are sensitive to collections, and to a lower level to rankers and performance measures;
    \item While QPP has potential applications such as selective query processing, the utility of current methods remains restricted due to their limited predictive power across diverse rankers;
    
\end{itemize}

Our study and our findings underscore the fragility of existing QPP approaches in handling rankers and their limited applicability for downstream tasks. They highlight the need for QPP metrics that can bridge the gap between sparse and dense retrieval paradigms while addressing the diversity of collections.

The remainder of this paper is organized as follows: Section~\ref{Sec:ExperimentalSetup} outlines the experimental setup, including datasets, rankers, and predictors. 
Section~\ref{Sec:Results} presents a detailed analysis of QPP performance across performance measures, rankers, and collections. Section~\ref{Sec:RobustnessDowstream} explores the downstream applications of prediction of the performance and selective query processing, and show their limitations. Section~\ref{Sec:RelatedWork} is the related work.  Section~\ref{Sec:Conclusion} concludes with a discussion of challenges, insights, and future directions.

\section{Experimental Setup and Evaluation Criteria}
\label{Sec:ExperimentalSetup}

\subsection{Data and Systems}

\begin{wraptable}{r}{0.6\textwidth} 
\centering
\small 
\renewcommand{\arraystretch}{0.9} 
\setlength{\tabcolsep}{3pt} 
\begin{tabular}{@{}p{1.7cm}|p{1.8cm}|p{1.2cm}|p{1cm}|p{1cm}@{}}
\textbf{Collection} & \textbf{Domain} & \textbf{\# D. (Q.)} & \textbf{\#W.} & \textbf{\#Qrel} \\
\hline
ROBUST & News, Gov., Legal & 522,006 (249) &  2.5 & 69.92  \\ 
GOV2 & Government documents & 25,177,288 (149) &  3 & 179.44 \\ 
WT10G & Web & 1,688,420 (99) & 4.2 & 59.79 \\ 
MS-MARCO DL 19 \& 20 & Web & 8,841,823 (97) & 6 & 107.15 \\ 
\end{tabular}
\caption{Data collections used in QPP experiments. The 3rd col. is the number of documents and queries in the collection; the 4th one (\# Words) is the average query length in terms of words; the last one (Qrel) is the average number of relevant documents per query.}
\label{tab:data_collections}
\end{wraptable}

We use four well-established TREC reference collections (See Table~\ref{tab:data_collections}); we use topic title as queries. 

On reference systems, we cover a spectrum of rankers: (a) Sparse Retrieval: BM25, DFree, (b) Dense Hybrid: SPLADE, and (c) Contextual Dense Retrieval: ColBERT. A summary of these systems is presented in Table~\ref{tab:reference_systems}, along with their strengths and weaknesses. A detailed description of their inner workings is provided below.

BM25 is a probabilistic ranker and part of the family of term-frequency-inverse-document-frequency (TF-IDF) based ranking functions~\cite{robertson2009probabilistic}. It scores documents based on the occurrence of query terms, adjusted by document length and term frequency. BM25 is acknowledged for its robustness and efficiency across a range of collections. For this reason, it is widely used as a baseline in IR.  Its parameters can be fine-tuned to better suit various datasets and query characteristics. 

DFree (Divergence from Randomness Model) \cite{amati2002probabilistic} estimates the informativeness of a term based on its deviation from a random distribution in documents. DFree captures more nuanced term significance by modeling randomness and term frequency distribution. DFree can handle cases where term significance varies greatly across documents, making it effective in domains with skewed or specialized vocabularies. This can provide more informative retrieval for collections with domain-specific language, which may benefit QPP accuracy in these contexts. DFree performance can be sensitive to collection characteristics, especially when applied to more general or heterogeneous datasets.

Both BM25 and DFree  are inherently sparse, meaning they rely on exact term matches between the query and document. This sparsity may limit their effectiveness on collections with complex queries or vocabulary mismatch, potentially impacting QPP, as predictions may not generalize well across queries.

SPLADE is a more recent ranker that combines sparse lexical representations with dense embeddings to capture both exact term matches and semantic meaning~\cite{formal2021splade}. By expanding queries and documents into sparse representations, it can handle vocabulary mismatch and subtle semantic similarities more effectively than  sparse models.
SPLADE’s hybrid approach enhances flexibility in capturing relevant information beyond exact term matches.

ColBERT~\footnote{\url{https://github.com/stanford-futuredata/ColBERT}} 
is a dense ranker that uses contextual embeddings from a pre-trained BERT model. It operates by embedding query and document terms independently, but with each term embedding contextualized by its surrounding words. ColBERT introduces a late interaction mechanism: instead of directly combining query and document embeddings into a single representation, it retains term-level embeddings, allowing the model to match query and document terms 
at retrieval time. 


\begin{table}[ht]
\centering
\renewcommand{\arraystretch}{1.4} 
\setlength{\tabcolsep}{8pt} 
\caption{Reference systems used in experiments}
\begin{tabular}{p{4cm}|p{9cm}} 
    \textbf{System} & \textbf{Strengths / Weaknesses} \\
    \hline
    BM25: Probabilistic model based on term frequency and document length normalization  
    & \textbf{Pro}: ROBUST baseline across collections; easy to tune.  
    \newline \textbf{Cons}: Limited to exact term matches, possible vocabulary mismatch. \\ 
    \hline
    DFree: Model estimating term informativeness by deviation from randomness  
    & \textbf{Pro}: Effective for specialized vocabularies.  
    \newline \textbf{Cons}: Limited to exact term matches; sensitive to collection characteristics. \\ 
    \hline
    SPLADE: Combines sparse lexical terms and dense embeddings for capturing both term matches and semantic meaning  
    & \textbf{Pro}: Effective for complex queries and semantically diverse data; robust in low-overlap vocabularies.  
    \newline \textbf{Cons}: High computational cost. \\ 
    \hline
    ColBERT: Dense ranker using contextual embeddings from BERT, with late interaction for term-level matching  
    & \textbf{Pro}: Effective for complex queries; balances semantic depth with efficiency, ranking relevant results close to the top.  
    \newline \textbf{Cons}: Computationally heavy. \\ 
\end{tabular}
\label{tab:reference_systems}
\end{table}

The ranker systems and their parameters were chosen based on previous studies. For BM25 and DFree, we used Terrier and we kept the default parameters set there~\cite{ounis2005terrier}, although the hyper-parameters could be fine-tuned to better suit each collection. 

For these two sparse models, we also used their variants when using automatic query expansion (QE). 
Bo2  is a pseudo-relevance feedback automatic QE method. It is designed to expand the original query with additional terms that are statistically likely to improve retrieval performance by capturing more relevant documents~\cite{amati2002probabilistic}. The Bo2 model scores candidate expansion terms using  
the probability of the term appearing in relevant documents versus its general frequency in the collection. We use this method for both sparse models BM25 and DFree. In the experiment part we used $5$ documents and $2$ terms for BM25 and with $20$ documents and  $5$ terms as there were the most effective when we varied these hyper-parameters.
 
For dense models, we used SPLADE v2 (Sparse Lexical and Expansion Model for Information Retrieval)\,\cite{formal2021splade}. The \texttt{distilSPLADE v2}\footnote{\url{https://github.com/naver/splade}} model improve performance through techniques such as hard-negative mining, distillation, and enhanced initialization of the pre-trained language model\,\cite{spladev2}. For ColBERT, we utilized the latest version, ColBERTv2\,\cite{ColBERTv2}, and used its checkpoint\,\footnote{\url{https://github.com/stanford-futuredata/ColBERT}} trained on the MS MARCO Passage Ranking dataset. We evaluated both SPLADE and ColBERT rankers using the BEIR\,\cite{thakur2021beir} framework to ensure a comprehensive performance assessment.

Query Performance Prediction (QPP) estimates retrieval systems effectiveness for a given query, offering valuable insights for search effectiveness and query processing. Despite extensive research,  QPPs face critical challenges in generalizing  across diverse retrieval paradigms and collections. This paper provides a comprehensive evaluation  of state-of-the-art QPPs (e.g., NQC, UQC), LETOR-based features, and newly explored dense-based predictors. Using diverse  sparse rankers (BM25, DFree without and with query expansion) and hybrid or dense (SPLADE and ColBert) rankers and diverse test collections—ROBUST, GOV2, WT10G, and MS MARCO—we investigate the relationships between predicted and actual performance, with a focus on generalization and robustness. Results show significant variability in  predictors accuracy, with collections as the main factor and rankers next.  Some sparse predictors perform somehow on some collections (TREC ROBUST and GOV2) but do not generalise to other collections (WT10G and MS-MARCO).   While some predictors show promise in specific scenarios, their overall limitations constrain their utility for applications. We show that QPP-driven selective query processing offers only marginal gains, emphasizing the need for improved predictors that generalize across collections, align with dense retrieval architectures and are useful for downstream applications. 
 We will publicly release our data and code following acceptance\footnote{\url{https://anonymous.4open.science/r/UncoveringTheLimitationsofQPP-346C/README.md}}.

\subsection{Predicting features and Predictive Models}
We employed three categories of features: state-of-the-art (SOTA) and LETOR features derived from sparse models, as well as features based on dense representations.

In the \textbf{SOTA} category, we consider  four features widely used in the QPP litterature:
\textit{Normalized Query Commitment (NQC)}~\cite{shtok2009predicting} measures the dispersion of retrieval scores for the top-ranked $k$ documents with normalization by query-specific corpus score\footnote{A score is obtained between the query and the corpus by treating the corpus as a single (large) document.}. It captures the intuition that a high-quality query tends to retrieve relevant documents with similar relevance scores, resulting in lower score variance, while low-quality queries show higher score variance among top documents due to increased retrieval noise.
\textit{Unnormalized Query Commitment (UCQ)}~\cite{shtok2012predicting} evaluates the variance in retrieval scores among the top-ranked $k$ documents, without normalizing by the query-specific corpus score. The UQC metric is based on the idea that a lower variance in scores among top-ranked documents indicates a more effective query, as it suggests consistent relevance among retrieved results, where as a higher variance may indicate a less effective query with more irrelevant results in the top ranks.
\textit{Weighted Information Gain (WIG)}~\cite{zhou2007query}  
    corresponds to the divergence between the mean of the top-ranked $k$ document scores and the mean of the entire set of document scores. The intuition behind WIG is that higher retrieval scores among the top ranked $k$ documents typically indicate a stronger match to the query, implying better query performance. Lower scores, on the contrary, suggest that the query may not align well with relevant documents, potentially indicating a lower-quality query.
 \textit{Query Feedback (QF)}~\cite{zhou2007query} uses pseudo-relevance feedback and estimates how robust a query might perform by examining the similarity between the original query and an expanded version generated from the top-ranked documents retrieved in an initial search. QF is estimated as the percentage of overlap at some rank $q$ between the returned document lists for the original query and the expanded query induced from the initial retrieved documents.
        

A \textbf{LETOR-based query feature} reflects a score on the matching of the query-document values at a certain rank $k$. Originally, LETOR query-document matching values were introduced in the context of Learning to rank~\cite{qin2010letor} where the goal is to learn a ranking function to better rank the top-retrieved documents given a query. Terrier IR~\cite{macdonald2013learning} implements 39 weighting models that can be used to estimate query-dependent document matching scores~\footnote{http://terrier.org/docs/current/javadoc/org/terrier/matching/models/package-summary.html}. 
LETOR features were used as query features~\cite{balasubramanian2010learning,chifu2018query}, they did not apply per-query normalization. To generate query features from LETOR features, \cite{balasubramanian2010learning,chifu2018query} just used different statistical summary functions (e.g., Mean, Std, Max) over the top-ranked $k$ query-document matching values for a given query. For LETOR features, the longer the query, the higher the feature value. Although this had no impact on previous application such as for the document re-ranking of documents, we found out that it has an impact when using LETOR features as  QPPs. We tried different LETOR  normalizations. When the normalization factor is uniform across the documents, the normalization can be done outside the summary functions~\cite{shtok2012predicting}. In this paper, we  normalize the LETOR features according to the query length (i.e., the number of effective query terms).



More formally, let $q$ be a query in the set of queries $\mathcal{Q}$ and $\mathbb{D}_{q,n}$ be the set of the top-ranked $k$ retrieved documents for  $q$. The $i$-th summarized LETOR feature, $\mathtt{SLF}_{i}(\mathbb{S},q)$ is calculated as follows:

\begin{equation}
\label{eq:SLF}
\mathtt{SLF}_i({\mathbb{S},q}) = \frac{1.0}{\mathcal{N}_{q}} \varphi_{\mathbb{S}} (\{\mbox{LF}_i(d,q)\}, q, \mathbb{D}_{q,k} ),
\end{equation}
\noindent where $d \in \mathbb{D}_{q,k}$ is a document, 
$\{\mbox{LF}_i(d,k)\}$ is the set of values for the LETOR feature $i$ for each couple (d, q), $\varphi_{\mathbb{S}}$ is a summary function with $\mathbb{S} \in \{Min, Max, Mean, Q_{1}, Median, Q_{3}, Std, Var, and\; Sum \}$, and ${\mathcal{N}_{q}}$ is the query-specific normalization factor. In this paper, we define the query-specific normalization factor as follows:
\begin{equation}
\label{eq:Normalization}
{\mathcal{N}_{q}} = \sum_{q_{j}}^{|q|} [tcf_{q_{j}}>0],
\end{equation}
where $q_{j}$ is a query term of the query $q$, $tcf_{q_{j}}$ is the corpus frequency of the query term $q_{j}$, $[tcf_{q_{j}}>0]$ = 1 when true, and 0 otherwise. In other words, the normalization factor $\mathcal{N}_{q}$ is defined as the count of effective query terms.

After a first analysis of the most correlated LETOR features, in the study we report here, we limit ourselves to the four following LETOR features: L.BM25, L.DFree, L.lemur, and L.In\_ExpC2 where each matching model is used with its default parameters as Terrier IR\footnote{http://terrier.org} and for which we consider the mean aggregation -also after a first analysis of the strongest correlations.



In the \textbf{Dense model} category, we include two embedding-based features from \cite{arabzadeh2021bert}, which utilize bi-encoder ($B_{bi}$) and cross-encoder ($B_{cross}$) architectures. Both rely on BERT's representational capacity to predict query-document relevance, though their approaches differ significantly.
\textit{$B_{bi}$} encodes queries and documents independently using a Siamese network architecture with two parallel BERT towers. The relevance between a query and a document is determined by computing their similarity, typically via cosine similarity or dot product. This design allows for pre-computing document embeddings, making it highly efficient and scalable for large-scale retrieval tasks.
\textit{$B_{cross}$}, on the other hand, processes the query and document together as a single input sequence by concatenating the query with the top-1 retrieved document. This unified BERT model captures detailed token-level interactions, enabling precise relevance predictions. However, it requires evaluating each query-document pair individually, resulting in higher computational costs.
$B_{bi}$ prioritizes efficiency and scalability, $B_{cross}$ focuses on precision; making the two complementary. The authors reported the correlation of these features with MRR@10 on the MS-MARCO datasets.


For predictions based on QPP models, we consider four\footnote{in our experiments, we consider more than that, but we decided to report some of them since the results are consistent across models} models:
\textit{Single-variable models}: Here we use a unique predictor; there is no training.\textit{ Multiple-variable models}: We use    
    Multiple Linear regression (LR), Random Forest (RF) and Support Vector Machine (SVM): In that cases we use cross-validation as explained in Section~\ref{Subsec:training}.

\subsection{Evaluation Measures}
\textbf{For evaluating the retrievers}—and, consequently, the values that QPP aims to predict—we focus on three commonly used effectiveness measures:
\textit{NDCG} balances relevance with rank position, rewarding rankers  that rank highly relevant documents closer to the top.
\textit{MAP} captures the overall retrieval quality across all relevant documents, providing a comprehensive view of performance.
\textit{P@10} is precision-oriented and emphasizes the top 10 documents, which is critical in user-facing applications where users typically do not look beyond the first page of results.
\textit{MRR@10} (Mean Reciprocal Rank)  evaluates the system's ability to place the first relevant document as high as possible in the ranking.  At rank 10, MRR@10 is calculated as the average reciprocal rank of the first relevant document for each query, but only considering the top 10 results. 


\textbf{To evaluate the efficiency of QPP}, we use also well-established measures from QPP literature:
    \textit{Correlation}: Both Pearson's $r$ and Kendall's $\tau$ correlation. While $r$  assume a linear correlation  $\tau$ does not; they are complementary. $\tau$ is more sensitive to ties. In the case of ties, $\tau$ may result in lower values of correlation because tied pairs are excluded from the calculation of concordant and discordant pairs. On the other hand, $\tau$ is robust to outliers but $r$ is not. In the case of predictors, we have both ties and outliers~\cite{chifu2024can}. 
    \textit{Machine learning}: we use the usual measures to evaluate regression models:  
    Mean Absolute Error (MAE), the average absolute difference between predicted and actual values, 
    Mean Squared Error (MSE) which penalizes  the larger errors more heavily than smaller errors, 
    Root Mean Squared Error (RMSE) which makes the MSE value to the original units. RMSE is highly affected by outlier values as it assumes that errors are unbiased and follow a normal distribution.  Median Absolute Error (MedAE) is like MAE, but uses the median over the mean and thus is less sensitive to outliers.
    R-squared (R$^2$)  measures how much variable in the target variable is explained by the machine learning model; it is calculated by dividing the variance of the predicted values by the variance of actual values; 1 meaning the regression perfectly fits the data. We report MAE, RMSE, MedAE and R$^2$
    \footnote{Formulas:  \url{https://www.appsilon.com/post/machine-learning-evaluation-metrics-regression}}.

\subsection{Training and Other Experiment Setup}
\label{Subsec:training}

In some cases, it is necessary to train models. To ensure the independence of the training and test sets while maintaining a sufficient number of queries in both, we employed two-fold cross-validation~\cite{wong2017dependency}. Specifically, the process is as follows: the set of queries is split into two equal parts—denoted as $Q_A$ and $Q_{\overline{A}}$. In the first fold, $Q_A$ is used for training, and $Q_{\overline{A}}$ is used for testing. In the second fold, the roles are reversed: $Q_{\overline{A}}$ is used for training, and $Q_A$ is used for testing.
To ensure comparability with methods that do not require a train/test split, we report the final results as the average of the results obtained on the two test sets. This approach ensures that all queries are used in the evaluation process. Additionally, the same query splits are applied consistently across all methods requiring training.
For statistical significance analysis, we use two-tailed paired t-test with Bonferroni correction ($^\ddagger$ means p-value $<$ 0.01 while $^\dagger$ is for $<$ 0.05). In theory there is no need to report the p-values as we know the number of observations. As for example, with $102$ queries (pairs of observation), the degree of freedom is $100$; and a difference of $0.195$ (resp. $0.254$) is statistically significant when considering a significance level of 0.05 for 2-tailed (resp. 0.01) - See Table at \url{https://www.statology.org/pearson-correlation-critical-values-table/}; we however report using $^\dagger$.


\section{Analysis and Results}
\label{Sec:Results}

\subsection{Feature effectiveness and limitations}
\label{subsec:feature}

\textbf{Comparison of features} We evaluate the effectiveness of the three types of features in terms of their individual relations with performance based on the correlation. Here we focus on NDCG performance measure (the sensitivity to the evaluation measure is further evaluated in Section~\ref{subsec:mesure_sensitivity}).  

\begin{table}[ht]
\centering
\caption{
Individual features - Correlation using Pearson $r$ and Kendall $\tau$ coefficients between NDCG performance measure and predicted values. Predictors include state-of-the-art predictors (top rows), summarized LETOR features calculated on BM25 run (middle rows, with mean aggregator), and BERT-based features (last rows). We consider BM25 and SPLADE systems (left and right parts of the table, respectively). $\dagger$ and $\ddagger$ indicate p-values $<0.05$ and $<0.01$, respectively. Queries are the topic titles.
}
\setlength{\tabcolsep}{5pt} 
\renewcommand{\arraystretch}{1.2} 
\resizebox{0.65\linewidth}{!}{ 
\begin{tabular}{l|cc|cc|cc|cc}
\toprule
NDCG & \multicolumn{4}{c|}{BM25} & \multicolumn{4}{c}{SPLADE} \\
& \multicolumn{2}{c|}{ROBUST} & \multicolumn{2}{c|}{MS-MC.} & \multicolumn{2}{c|}{ROBUST} & \multicolumn{2}{c}{MS-MC.} \\
& $r$ & $\tau$ & $r$ & $\tau$ & $r$ & $\tau$ & $r$ & $\tau$ \\
\midrule
UQC & .407$^\ddagger$ & .322$^\ddagger$ & -.123 & -.025 & .439$^\ddagger$ & .328$^\ddagger$ & .401$^\ddagger$ & .277$^\ddagger$ \\
NQC & .354$^\ddagger$ & .285$^\ddagger$ & -.010 & -.005 & .295$^\ddagger$ & .226$^\ddagger$ & .212$^\dagger$ & .156$^\dagger$ \\
WIG & .342$^\ddagger$ & .236$^\ddagger$ & .027 & -.080 & .354$^\ddagger$ & .161$^\ddagger$ & .179 & .086 \\
QF  & .394$^\ddagger$ & .265$^\ddagger$ & .146 & .106 & .436$^\ddagger$ & .297$^\ddagger$ & .418$^\ddagger$ & .320$^\ddagger$ \\
\midrule
L.BM25 & .459$^\ddagger$ & .321$^\ddagger$ & .149 & .120 & .279$^\ddagger$ & .209$^\ddagger$ & -.116 & -.104 \\
L.DFree & .443$^\ddagger$ & .290$^\ddagger$ & .157 & .116 & .268$^\ddagger$ & .218$^\ddagger$ & -.083 & -.088 \\
L.Lemur & .456$^\ddagger$ & .326$^\ddagger$ & .121 & .103 & .200$^\dagger$ & .163$^\ddagger$ & -.106 & -.094 \\
L.InExp2 & .424$^\ddagger$ & .327$^\ddagger$ & .070 & .100 & .283$^\ddagger$ & .228$^\ddagger$ & -.109 & -.101 \\
\midrule
B$_{bi}$ & .151$^\dagger$ & .104$^\dagger$ & -.166$^\dagger$ & .157$^\ddagger$ & .122 & .082 & .204$^\dagger$ & -.078 \\
B$_{cross}$ & .069 & .034 & .009 & -.042 & .032 & .019 & .006 & .111 \\
\bottomrule
\end{tabular}
} 
\label{tab:NDCG_Rho_Tau}
\end{table}

\footnotetext{For LETOR features, we used the ones calculated on BM25 runs since they are calculated in Terrier and corresponds to sparse features. We did the same for SOTA sparse features.}

In Table~\ref{tab:NDCG_Rho_Tau}, Pearson correlation ($r$) values are consistently higher than Kendall ($\tau$) values across all predictors, reflecting stronger linear relationships than rank consistency. Indeed, $r$ measures the linear relationship between two variables while $\tau$ measures the rank-based association between two variables. It assesses how consistently the relative order (ranking) of the values is preserved between the two variables. A lower $\tau$  may imply that while the two variables may be linearly related (reflected by the $r$), their ranks or orderings are not perfectly aligned. On the other hand, $\tau$ is less sensitive to extreme values (outliers) than $r$  but more sensitive to small changes in ranks and ties. 

On ROBUST collection, the LETOR features (L.BM25, L.DFree, L.Lemur, L.InExp2) generally outperform the SOTA QPP metrics (NQC, UQC, WIG, QF) for both BM25 but it is the other way around for SPLADE ranker; the correlation values are still relatively weak for both BM25 and SPLADE. Considering BM25 ranker, LETOR features exhibit slightly better predictive power compared to SOTA predictors, but the correlation values remain modest, indicating room for improvement in capturing query performance even in sparse retrieval contexts. Dense-based features have the weakest correlations. For SPLADE, correlation values for LETOR features are lower than for BM25, suggesting that these features are less effective in dense retrieval contexts.

BERT-based predictors are very weakly or not correlated to the actual NDCG.

\begin{wrapfigure}{r}{0.65\textwidth}  
    \centering
    \includegraphics[width=0.5\textwidth]{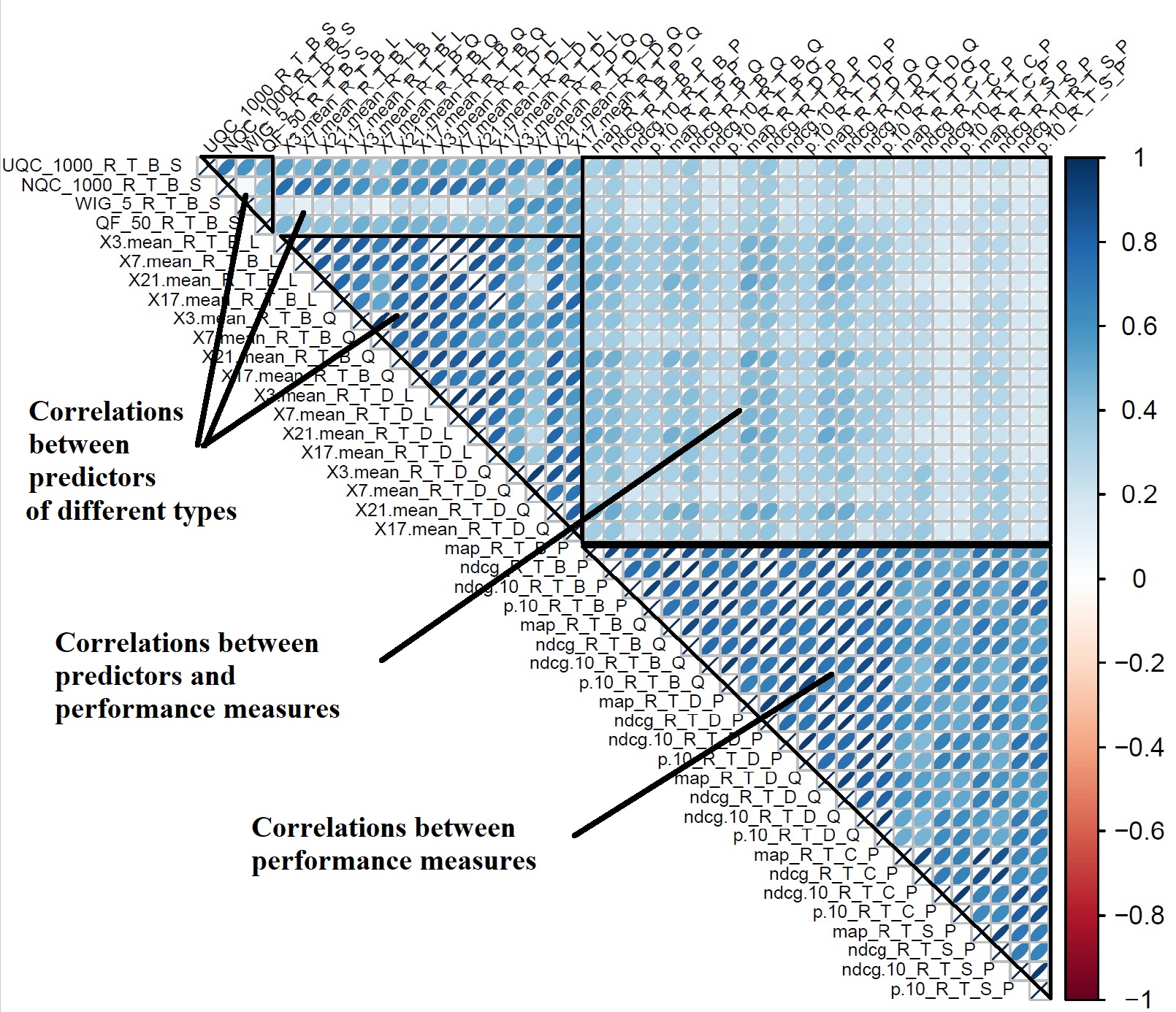}
          \caption{ROBUST collection - $r$ correlation. The strongest correlations are among performance measures; correlations between predictors and performance measures are weak. }
    \label{fig:RobustCorrelation}
\end{wrapfigure}

The results do not generalize well on MS-MARCO collection where none of the predictors is correlated with NDCG for BM25 ranker but some are weakly correlated for SPLADE.

NDCG correlations provide insights into general effectiveness, but further analysis with metrics like P@10 or MAP may reveal additional nuances in predictor behaviour (See Section~\ref{Subsec:Collection_Sensitivity} for other performance measures).

Both SOTA and LETOR features appear  more accurate than  BERT-based ones, neither approach provides particularly strong correlation with actual performances -this is also confirmed when plotting the link between predictions and actual values (See some examples in Figure~\ref{fig:teaser} where the ranker is BM25 on ROBUST). 
These results point to the potential need for new predictors tailored to be robust across model types (sparse/dense) and across collections.

In Figure~\ref{fig:RobustCorrelation} we can see the $r$ correlations between performance measures and QPPs. 
The top right square shows only weak correlations; they correspond to correlation between performance measures and predictors. The right bottom triangle corresponds to the correlations between performance measures. 

The correlations are much higher. On the left are displayed the correlations among QPPs: the very left-side triangle is between SOTA QPPs, the other triangle is among LETOR features. We can see that SOTA QPPs are not much correlated among them, contrarly to LETOR ones. 

The left-side rectangle is between SOTA and LETOR features; they are not much correlated. All in all the weaker correlations are between QPPs and  performance measures. 
Figure~\ref{fig:RobustCorrelation} reports the results for ROBUST collection, but the same holds for the other collections.

\textbf{Feature interaction effects}: a single feature might not be enough for prediction. Here we compare notable interaction effects among features using multiple features. We consider  the $Q_A$ and $Q_{\overline{A}}$ framework explained in Section~\ref{Subsec:training}. We report the results for Linear Regression and Random Forest. First we combine the SOTA features, then the LETOR features, and both. We did not consider BERT-based features considering their weak correlation with actual measures. 

Table~\ref{tab:FeatureModel} reports the results on ROBUST collection for NDCG-We use the ones calculated on BM25 ranking (best). The correlations are slightly larger than when using single features (See Table~\ref{tab:NDCG_Rho_Tau}). The correlations however are above $0.5$.



\begin{table}[ht]
\centering
\caption{
Combined features: Correlation between QPPs and NDCG on ROBUST using BM25 and SPLADE. The 4 SOTA features combined (first two rows), the 4 LETOR combined (second series), and all 8 QPPs combined (last two rows).
}
\setlength{\tabcolsep}{4pt} 
\renewcommand{\arraystretch}{1.2} 
\resizebox{0.7\textwidth}{!}{ 
\begin{tabular}{l|c|cc|cc|cc|cc}
\toprule
NDCG & Model & \multicolumn{4}{c|}{BM25} & \multicolumn{4}{c}{SPLADE} \\
& & \multicolumn{2}{c|}{ROBUST} & \multicolumn{2}{c|}{MS-MC.} & \multicolumn{2}{c|}{ROBUST} & \multicolumn{2}{c}{MS-MC.} \\
& & $r$ & $\tau$ & $r$ & $\tau$ & $r$ & $\tau$ & $r$ & $\tau$ \\
\midrule
SOTA & LR & .482$^\ddagger$ & .347$^\ddagger$ & .364$^\ddagger$ & .307$^\ddagger$ & 0.527$^\ddagger$ & 0.374$^\ddagger$ & .237$^\dagger$ & 0.194$^\ddagger$ \\ 
- & RF & .459$^\ddagger$ & .316$^\ddagger$ & .327 & .248$^\ddagger$ & 0.494$^\ddagger$ & 0.324$^\ddagger$ & 0.344$^\ddagger$ & 0.247$^\ddagger$ \\ 
\midrule
LETOR & LR & .424$^\ddagger$ & .297$^\ddagger$ & -.060 & -.029 & .498$^\ddagger$ & .362$^\ddagger$ & .128 & .040 \\
- & RF & .400$^\ddagger$ & .263$^\ddagger$ & .119 & .069 & .504$^\ddagger$ & .346$^\ddagger$ & .054 & .033 \\
\midrule
BOTH & LR & .449$^\ddagger$ & .328$^\ddagger$ & .067 & .037 & .499$^\ddagger$ & .362$^\ddagger$ & .333$^\ddagger$ & .227$^\ddagger$ \\
- & RF & .491$^\ddagger$ & .332$^\ddagger$ & .077 & .034 & .504$^\ddagger$ & .346$^\ddagger$ & .352$^\ddagger$ & .253$^\ddagger$ \\
\bottomrule
\end{tabular}
} 
\label{tab:FeatureModel}
\end{table}

\subsection{Sensitivity to the Evaluation Measure }
\label{subsec:mesure_sensitivity}

\textbf{Performance variability across effectiveness metrics}: In the previous results, we focus on  NDCG, here we analyse further the  QPP performance differences across evaluation metrics.

\begin{table}[ht]
\centering
\caption{
Individual features - Sensitivity to the evaluation measure - ROBUST collection and $r$ correlation.
}
\setlength{\tabcolsep}{4pt}  
\renewcommand{\arraystretch}{1.2} 
\resizebox{0.65\textwidth}{!}{ 
\begin{tabular}{l|cccc|cccc}
$r$ & \multicolumn{4}{c|}{BM25} & \multicolumn{4}{c}{SPLADE} \\
& NDCG & MAP & P@10 & MRR@10 & NDCG & MAP & P@10 & MRR@10 \\
\midrule
UQC & .407$^\ddagger$ & .336$^\ddagger$ & .230$^\ddagger$ & .221$^\ddagger$ 
& .439$^\ddagger$ & .325$^\ddagger$ & .219$^\ddagger$ & .143$^\dagger$  \\
NQC & .354$^\ddagger$ & .289$^\ddagger$ & .243$^\ddagger$ & .177$^\ddagger$ 
& .295$^\ddagger$ & .200$^\ddagger$ & .127$^\dagger$ & .066  \\
WIG & .342$^\ddagger$ & .286$^\ddagger$ & .199$^\ddagger$ & .219$^\ddagger$ 
& .354$^\ddagger$ & .317$^\ddagger$ & .257$^\ddagger$ & .226$^\ddagger$ \\
QF  & .394$^\ddagger$ & .326$^\ddagger$ & .285$^\ddagger$ & .250$^\ddagger$ 
& .436$^\ddagger$ & .413$^\ddagger$ & .378$^\ddagger$ & .250$^\dagger$ \\
\midrule
L.BM25   & .459$^\ddagger$ & .446$^\ddagger$ & .358$^\ddagger$ & .276$^\ddagger$ 
& .279$^\ddagger$ & .288$^\ddagger$ & .166$^\ddagger$ & .157$^\dagger$ \\
L.Dfree  & .443$^\ddagger$ & .432$^\ddagger$ & .343$^\ddagger$ & .262$^\ddagger$ 
& .268$^\ddagger$ & .277$^\ddagger$ & .164$^\ddagger$ & .139$^\dagger$ \\
L.Lemur  & .456$^\ddagger$ & .500$^\ddagger$ & .341$^\ddagger$ & .288$^\ddagger$ 
& .200$^\dagger$ & .245$^\ddagger$ & .141$^\dagger$ & .147$^\dagger$ \\
L.InExp2 & .424$^\ddagger$ & .421$^\ddagger$ & .325$^\ddagger$ & .227$^\ddagger$ 
& .283$^\ddagger$ & .288$^\ddagger$ & .151$^\dagger$ & .164$^\ddagger$ \\
\midrule
B$_{bi}$ & .151$^\dagger$ & .161$^\ddagger$ & .087 & .008 
& .122 & .089 & .119 & .086  \\
B$_{cross}$ & .069 & .066 & .022 & .045 
& .032 & .026 & -.051 & .086 \\
\end{tabular}
} 
\label{tab:PerformanceSensitivity_IndividualFeatures}
\end{table}

Table~\ref{tab:PerformanceSensitivity_IndividualFeatures} presents the Pearson correlation $r$ values for the QPP features across four evaluation metrics (NDCG, MAP, P@10, MRR@10) for two rankers (BM25 and SPLADE).

P@10 and MRR@10 demonstrate the lowest correlation values among the three metrics, likely because predicting top-10 precision is more sensitive to query-specific challenges and document rankings. NDCG and MAP correlations with the predicted value are closer one to the other. Actual MAP and NDCG values are also more correlated one to the other than with P@10 and MRR@10 (See Section~\ref{subsec:feature} - Figure~\ref{fig:RobustCorrelation}).

\begin{table}[h] 
\centering
\caption{Combined features - Evaluation measure sensitivity - ROBUST collection - $r$ correlation (notations as in Table 5).}
\setlength{\tabcolsep}{4pt} 
\renewcommand{\arraystretch}{1.1} 
\begin{tabular}{l|c|ccc|ccc}
\toprule
$r$ & Model & \multicolumn{3}{c|}{BM25} & \multicolumn{3}{c}{SPLADE} \\
& & NDCG & MAP & P@10 & NDCG & MAP & P@10  \\
\midrule
SOTA & LR & .482$^\ddagger$ & .392$^\ddagger$ & .316$^\ddagger$ & .527$^\ddagger$ & .454$^\ddagger$ & .381$^\ddagger$\\
 - & RF & .459$^\ddagger$ & .358$^\ddagger$ & .289$^\ddagger$ & .494$^\ddagger$ & .406$^\ddagger$ & .328$^\ddagger$\\
\midrule
LETOR & LR & .424$^\ddagger$ & .489$^\ddagger$ & .326$^\ddagger$ & .303$^\ddagger$ & .266$^\ddagger$ & .054\\
- & RF & .400$^\ddagger$ & .382$^\ddagger$ & .298$^\ddagger$ & .224$^\ddagger$ & .208$^\ddagger$ & .124$^\dagger$\\
\midrule
BOTH & LR & .449$^\ddagger$ & .478$^\ddagger$ & .273$^\ddagger$ & .498$^\ddagger$ & .439$^\ddagger$ & .340$^\ddagger$\\
- & RF & .491$^\ddagger$ & .440$^\ddagger$ & .356$^\ddagger$ & .504$^\ddagger$ & .428$^\ddagger$ & .336$^\ddagger$\\
\bottomrule
\end{tabular}
\label{tab:PerformanceSensitivity_Models}
\end{table}

The highest correlation is for NDCG;  indicating that these features are relatively better at predicting position-based relevance performance.  
The values are lower for MAP than NDCG, especially for SPLADE, where MAP shows weaker correlations overall. For SPLADE, correlations are generally weaker than for BM25 ranker, suggesting that traditional predictors struggle with dense retrieval settings. BERT-based are the weakest consistently accross the performance measures.

The effectiveness of QPP predictors  depends on the evaluation measure, with NDCG being the most aligned with their design in almost all the cases. 



For predictions based on combined features (See~Table~\ref{tab:PerformanceSensitivity_Models}), the correlation with performance measures is again better with NDCG than with the other performance measures, the worst being with P@10.

\subsection{Sensitivity to the collection}
\label{Subsec:Collection_Sensitivity}

\begin{table}[ht]
\caption{
Sensitivity to the collection. ** Due to a lack of computational resources, we were unable to obtain results for this collection using the SPLADE. }
\centering
\setlength{\tabcolsep}{2pt} 
\renewcommand{\arraystretch}{0.9} 
\begin{tabular}{l|llll|llll}
NDCG& \multicolumn{4}{c|}{BM25} & \multicolumn{3}{c}{SPLADE} \\
$r$& \footnotesize{ROBUST} & \footnotesize{GOV2} & \footnotesize{WT10G} & \footnotesize{MS-M.} &\footnotesize{ ROBUST} & \footnotesize{GOV2} & \footnotesize{WT10G} & \footnotesize{MS-M.}  \\
\midrule
UQC & .407$^\ddagger$ & .382$^\ddagger$ & .181 & -.123 &.439$^\ddagger$ &**&.271$^\ddagger$& .401$^\ddagger$     \\
NQC & .354$^\ddagger$ & .380$^\ddagger$ & .067 & -.010 &.295$^\ddagger$ &**&.108&.212$^\dagger$        \\
WIG & .342$^\ddagger$ & .315$^\ddagger$ & .125 & -.080 &.354$^\ddagger$ &**&.138&.179        \\
QF  & .394$^\ddagger$ & .237$^\ddagger$ & .161 & .146  &.436$^\ddagger$ &**&.416$^\ddagger$&.418$^\ddagger$        \\
\midrule
L.BM25   & .459$^\ddagger$ & .326$^\ddagger$& .205$^\dagger$  & .149 & .279$^\ddagger$   &&.041&-.116        \\
L.Dfree  & .443$^\ddagger$ & .347$^\ddagger$& .091 & .157 & .268$^\ddagger$   &**&.039&-.083        \\
L.Lemur  & .456$^\ddagger$ & .223$^\ddagger$& .146 & .121 & .200$^\dagger$   &**&.050&-.106        \\
L.InExp2 & .424$^\ddagger$ & .451$^\ddagger$& .231$^\dagger$  & .070 & .283$^\ddagger$   &&.094&-.109        \\
\midrule
B$_{bi}$ & .151$^\dagger$ & .136 &       &-.166$^\dagger$       &.122  &**  &   &   .204$^\dagger$    \\
B$_{cross}$ & .069 & -.046 &              &  .009 &.032 & **  &   &  .006  \\
\end{tabular}

\label{tab:CollectionSensitivity_IndividualFeatures}
\end{table}

The individual predictors do not generalize well across collections. Both SOTA and LETOR features behave closely for  ROBUST and GOV2, but stuck on WT10G or MS-MARCO. BERT$_{Bi}$ is slightly more correlated with NDCG on MS-MARCO for SPLADE confirming the design is quite specific. On SPLADE none of the positive results observed on ROBUST generalize.

\subsection{Sensitivity to the ranker }
\label{subsec:Ranking_Sensitivity}

\begin{table}[ht]
\caption{Sensitivity to the ranker. ROBUST collection}
\centering
\begin{tabular}{l|cc|cc|c|c}
$r$& \multicolumn{2}{c|}{BM25} & \multicolumn{2}{c}{DFRee} & \multicolumn{2}{c}{Other} \\
NDCG & -  & QE & - & QE & SPL. & Col.   \\
\midrule
UQC &  .407$^\ddagger$ &.367 & .340$^\ddagger$ & .371$^\ddagger$ & .439$^\ddagger$ & .244$^\ddagger$\\
NQC &  .354$^\ddagger$& .299 & .356$^\ddagger$ & .338$^\ddagger$ & .295$^\ddagger$ & .163$^\ddagger$\\
WIG &  .342$^\ddagger$& .440 & .346$^\ddagger$ & .318$^\ddagger$ & .354$^\ddagger$ & .246$^\ddagger$\\
QF  &  .394$^\ddagger$& .384 & .407$^\ddagger$ & .387$^\ddagger$ & .436$^\ddagger$ & .225$^\ddagger$\\
\midrule
L.BM25   & .459$^\ddagger$ & .367$^\ddagger$ &.465$^\ddagger$ &.419$^\ddagger$ & .279$^\ddagger$ & .188$^\ddagger$      \\
L.Dfree  & .443$^\ddagger$ & .299$^\ddagger$ &.454$^\ddagger$ &.340$^\ddagger$ & .268$^\ddagger$ &    .247$^\ddagger$   \\
L.Lemur  & .456$^\ddagger$ & .440$^\ddagger$ &..473$^\ddagger$ &.493$^\ddagger$ & .200$^\dagger$  &   .161$^\dagger$    \\
L.InExp2 & .424$^\ddagger$ & .384$^\ddagger$ &.427$^\ddagger$ &.376$^\ddagger$ & .283$^\ddagger$ &    .271$^\ddagger$   \\
\midrule
\end{tabular}
\label{tab:RankerSensitivity_IndividualFeatures}
\end{table}


We analyse QPP performance across sparse rankers (BM25, DFree) and more dense models  SPLADE and ColBert. We also compare the results with and without  QE. We report on two collections (ROBUST and MS-MARCO). Table~\ref{tab:RankerSensitivity_IndividualFeatures} shows that the both SOTA and LETOR \footnote{SOTA features are  calculated from the BM25 reference system; while LETOR are calculated based on the considered ranker retrieved documents.} predictors generalize well across rankers.

\section{Robustness and Downstream Impact}
\label{Sec:RobustnessDowstream}

\subsection{ANOVA and Main components}

We performed ANOVA to assess whether the correlation values differ significantly between rankers or collections. 

The total variation (Sum Sq) in correlation explained by the collection factor  is much higher than the one explained by the ranker factor (Table~\ref{tab:AnovaTable}). The same holds for the average variation.
The unexplained variation per degree of freedom for residuals is about 0.011 in both cases. Residuals represent the variation that is not explained by either Collection or Ranker factor. All in all, these results indicate that the collection factor explains a large portion of the variation, leaving little unexplained. 

\begin{table}[ht]
\caption{Analysis of Variance (ANOVA) Results}
\centering
\begin{tabular}{lrrrrr}
\hline
\textbf{Source} & \textbf{Df} & \textbf{Sum Sq} & \textbf{Mean Sq} & \textbf{F value} & \textbf{Pr(>F)} \\ 
\hline
Collection & 3   & 4.554 & 1.518 & 137.6 & $<$2e-16 \\ 
Residuals  & 604 & 6.662 & 0.011 &       &              \\ 
\hline
Ranker     & 5   & 0.275  & 0.05491 & 3.021  & 0.0106 *** \\ 
Residuals  & 602 & 10.941 & 0.01818 &       &              \\ 
\end{tabular}
\label{tab:AnovaTable}
\end{table}

\begin{figure}[ht]
    \centering
    \caption{
        Distributions of correlation values ($r$) between the performance metrics and the predictors are much affected by the collections than by the rankers. Distribution per ranker (left) and per collection (right).
    }
    
    \begin{subfigure}[t]{0.49\textwidth}
        \centering
        \includegraphics[width=\textwidth]{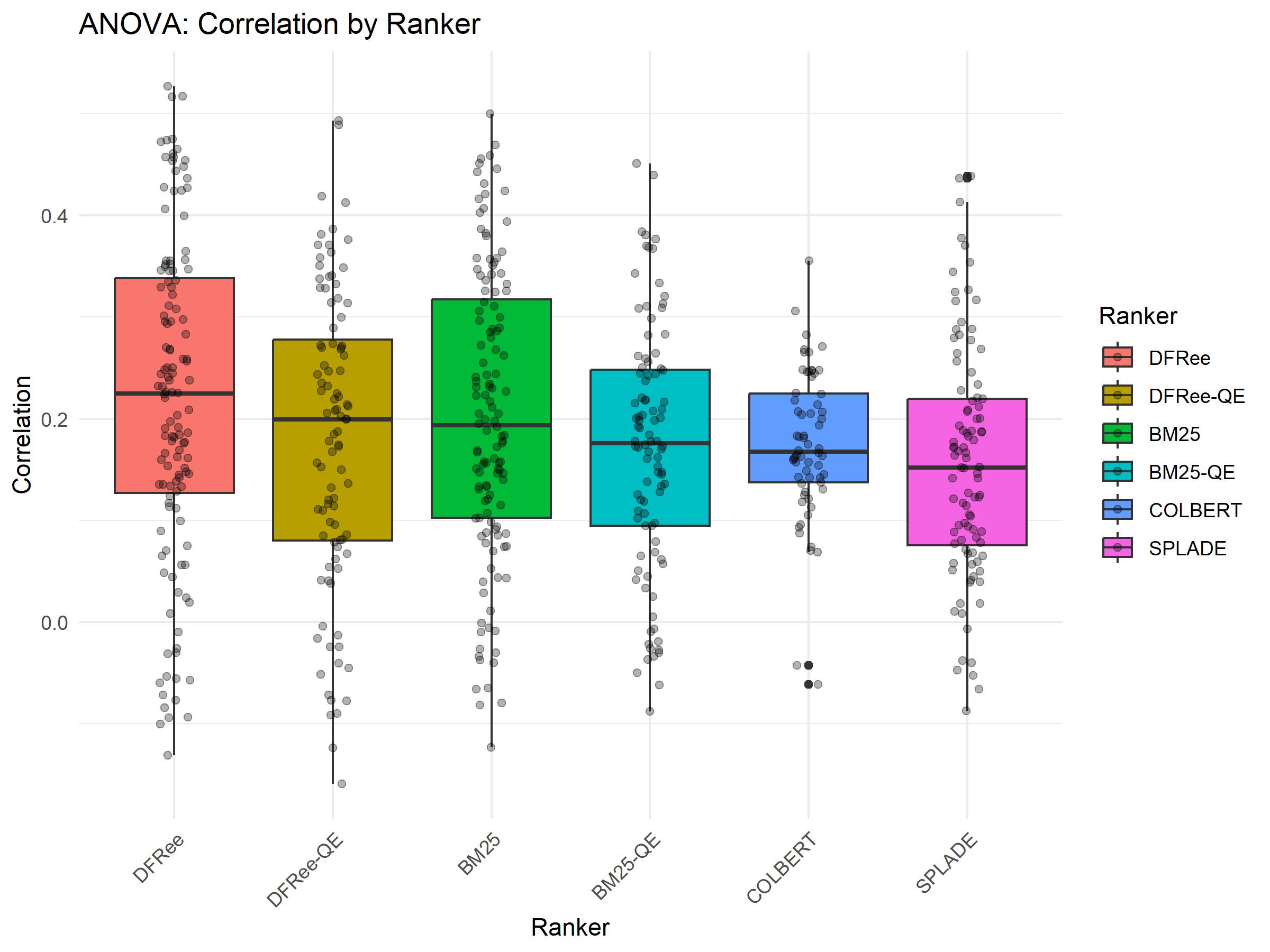}
        \caption{Per Ranker}
    \end{subfigure}
    \hfill
    \begin{subfigure}[t]{0.49\textwidth}
        \centering
        \includegraphics[width=\textwidth]{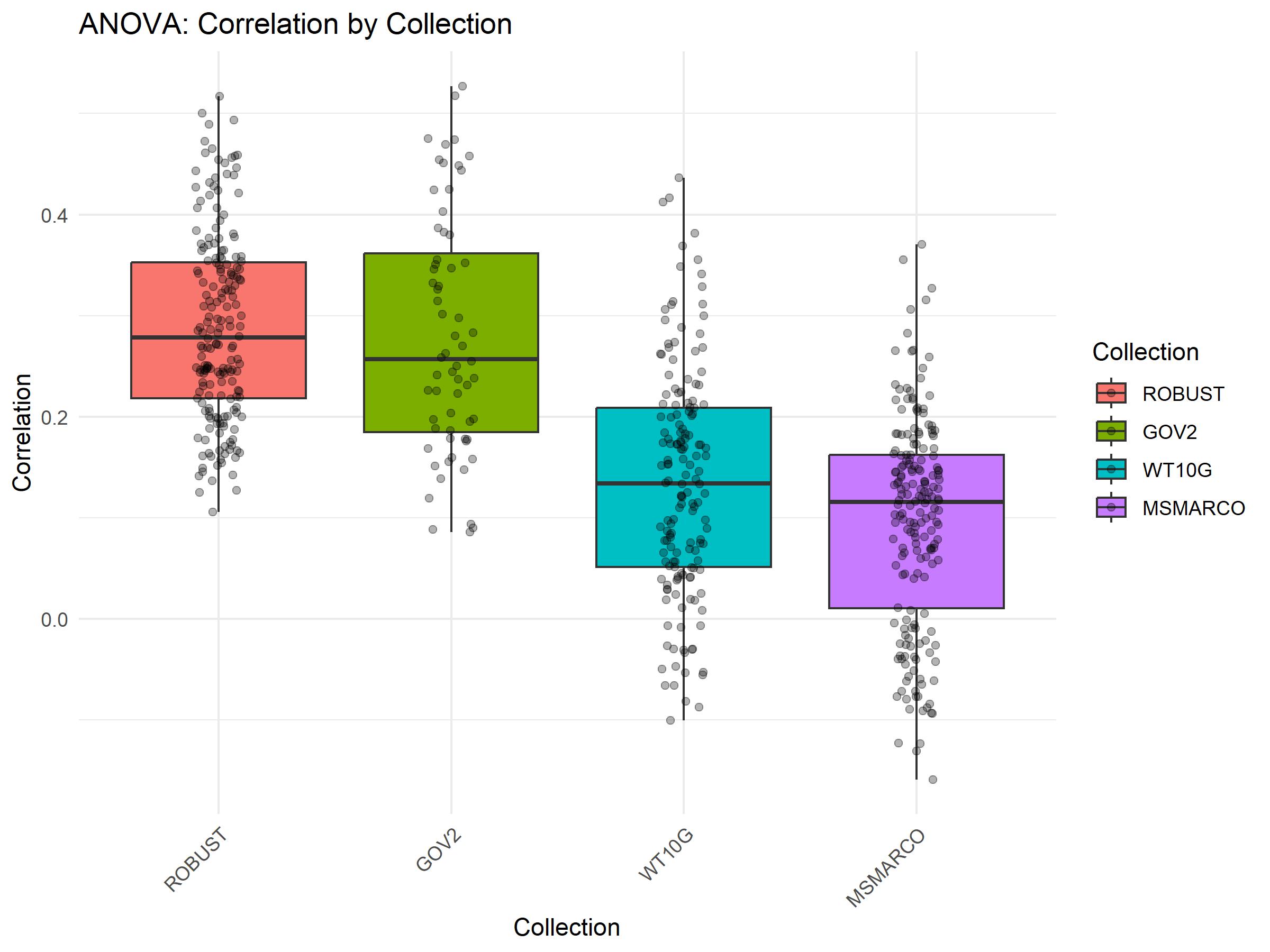}
        \caption{Per Collection}
    \end{subfigure}

    \label{fig:Anova}
\end{figure}

Boxplots in Figure~\ref{fig:Anova} visually represents the distributions of correlation values ($r$) between the performance metrics and the predictors, grouped by  (a) ranker and (b) collection. The boxplots were sorted by the median correlation value in decreasing order for both rankers and collections. These boxplots provide insight into the variability and central tendency of the correlations.

 From Figure~\ref{fig:Anova}, we can see that predictors are generally more correlated with performance measures on  ROBUST and MS-MARCO than on the two other collections (Sub-Figure (a)) and that there are slightly better correlated for DFRee ranker than for the other rankers (Sub-Figure (b)). The spread of the correlation values in each boxplot provides additional information about the consistency and reliability of the predictors' performance across different datasets and evaluation metrics. For example, the spread is lower for  ROBUST collection and for ColBert ranker. 

 Finally, we can see that the collection factor is much more important than the ranker factor, which is corroborated by the results of the ANOVA.

\section{Downstream Applications}
\subsection{Predicting the query performance}
One straightforward application is indeed to predict the performances. While Tables \ref{tab:FeatureModel} and \ref{tab:PerformanceSensitivity_Models} report the correlations between the performance measurements and the predicted values, here we report the error the system makes in the prediction.
From Table~\ref{tab:FeatureModelMAE} we can conclude that the prediction is not very accurate as we expected considering the weak correlations. For example the Mean Average Error of $0.167$ is to be put into the perspective of the value that is it supposed to predict. For example, $0.511$ in average NDCG for BM25 for example (See Table~\ref{tab:PerformanceRankers}). That means that the prediction will be $0.511 \pm 0.167$. Although MedAE is less sensitive to outliers, the error remains large.


\begin{table}[ht]
\centering
\caption{Performance for the different collections for different performance measures. ** could not be reported because of lack of computational resources.}
\small 
\setlength{\tabcolsep}{2pt} 
\renewcommand{\arraystretch}{0.9} 
\begin{tabular}{@{}l|llll|llll@{}}
& \multicolumn{4}{c|}{NDCG} & \multicolumn{4}{c}{MAP} \\
& \footnotesize{ROBUST} & \footnotesize{GOV2} & \footnotesize{WT10G} & \footnotesize{MS-M.} & \footnotesize{ROBUST} & \footnotesize{GOV2} & \footnotesize{WT10G} & \footnotesize{MS-M.}  \\
\midrule
BM25  & .511 & .556 & .435 & .576 & .240 & .272 & .174 & .336\\
BM25 QE &.532 & ** & .454 & .591& .260 & ** & .195 & .365\\
DFRee &  .523 & .582 & .458 & .575 &  .252 & .292 & .188 & .340 \\
DFRee QE & .551 & ** & .452 & .598 & .278 & ** & .189 & .373 \\
SPLADE & .452 & .405 & .376 & .744& .207 & .163 & .155 & .376\\
ColBert & .430 & **& .345& .767& .196& **& .131&.552 \\
\midrule
\end{tabular}
\label{tab:PerformanceRankers}
\end{table}

\begin{table}[ht]
\caption{Predictive model accuracy - TREC ROBUST - NDCG. }
\centering
\small 
\setlength{\tabcolsep}{2pt} 
\begin{tabular}{l|l|cc|cc|cc|cc}
 & NDCG & \multicolumn{4}{c|}{BM25} & \multicolumn{4}{c}{SPLADE} \\
 \hline
 & RF & MAE & RMSE & MedAE & $R^2$ & MAE & RMSE & MedAE & $R^2$\\
 & SOTA & .161 & .199 & .145 & -1.571 & .1445  & .181 & .121 & -1.175\\
 & LETOR & .172 & .206 & .158 & -2.640 & .169&.209&.144&-3.551\\
 & BOTH & .167 & .205 & .144 & -1.504 & .147&.179&.131&-1.473\\
\end{tabular}

\label{tab:FeatureModelMAE}
\end{table}

\subsection{Selective ranker}

We investigate the role of QPPs in guiding decisions like selectively applying query expansion  or choosing between sparse and dense rankers to optimize  retrieval effectiveness. Based on the findings of previous sections, we hypothesize that QPPs may be able to assist in identifying the most appropriate ranker 
for a  query. 

Specifically, we use the predicted performance value to guide the ranker choice. For example, if the predictor is positively correlated with the performance of a ranker without QE, a threshold can be defined on the predictor. Queries with predicted performance below this threshold would benefit from QE, while those above it can be processed without it. This selective strategy ensures that QE is applied only when it is expected to improve retrieval effectiveness, leveraging the predictive power of QPPs to dynamically optimize the ranking process. The same type de decision can be used to choose between sparse and dense rankers.
This not only optimizes efficiency by avoiding unnecessary computational overhead, but can enhances effectiveness by aligning the ranker with the specific query requirements.

Let $q$ be a query, $P(q)$ the predicted performance value for $q$ using a QPP, $T$ a predefined threshold on the predicted performance, $R_{\text{R1}}(Q)$ the ranker $R1$, $R_{\text{R2}}(q)$ the ranker $R2$, and $R_{\text{opt}}(q)$ the optimal ranker selected for $q$, defined  as follows:
\begin{equation}
\label{eq:Selective1}
R_{\text{opt}}(q) =
\begin{cases} 
R_{\text{R1}}(q), & \text{if } P(q) \leq T, \\
R_{\text{R2}}(q), & \text{if } P(q) > T.
\end{cases}
\end{equation}

\begin{figure}[ht]
    \centering
    \begin{subfigure}[t]{0.49\columnwidth} 
        \centering
         \includegraphics[width=\textwidth]{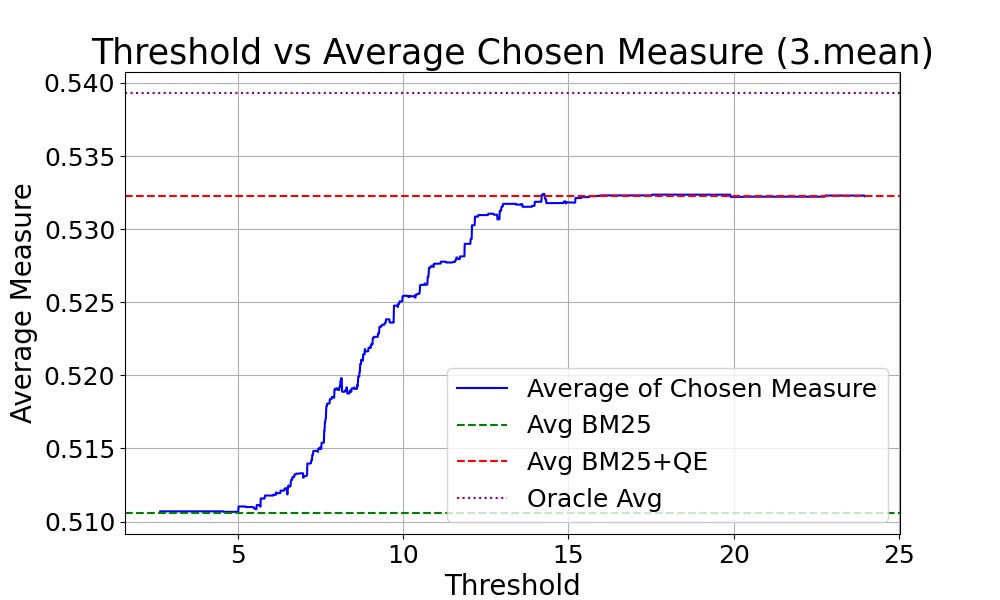}
        \caption{BM25 vs BM25 QE - Based on Letor feature L.BM25}
        \label{fig:subfig1}
    \end{subfigure}
    \hspace{-1mm} 
    \begin{subfigure}[t]{0.49\columnwidth} 
        \centering
        \includegraphics[width=\textwidth]{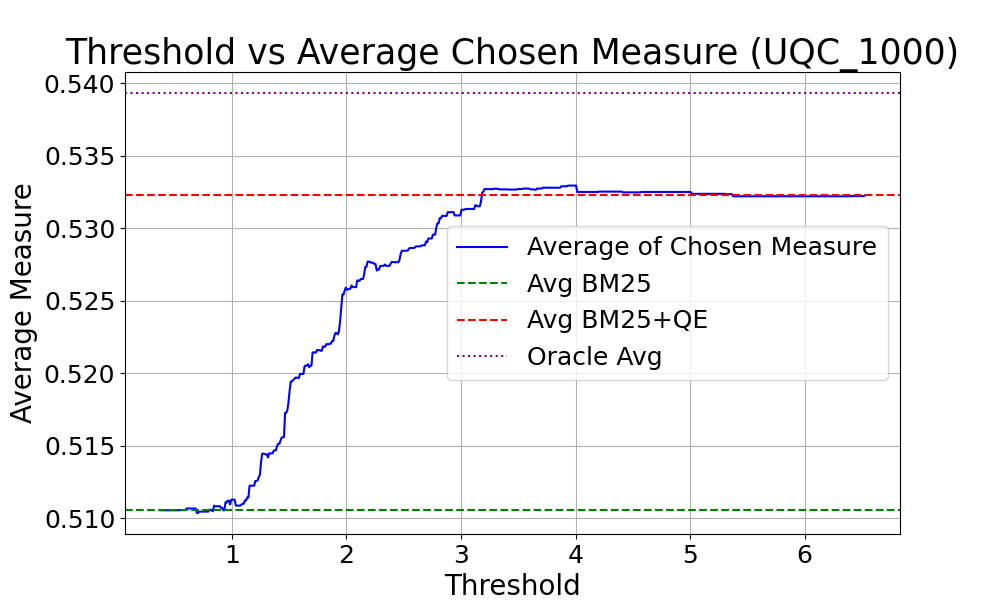} 
        \caption{BM25 vs BM25 QE - Based on SOTA UQC} 
        \label{fig:subfig2}
    \end{subfigure}
    \\
     \begin{subfigure}[t]{0.49\columnwidth} 
        \centering
         \includegraphics[width=\textwidth]{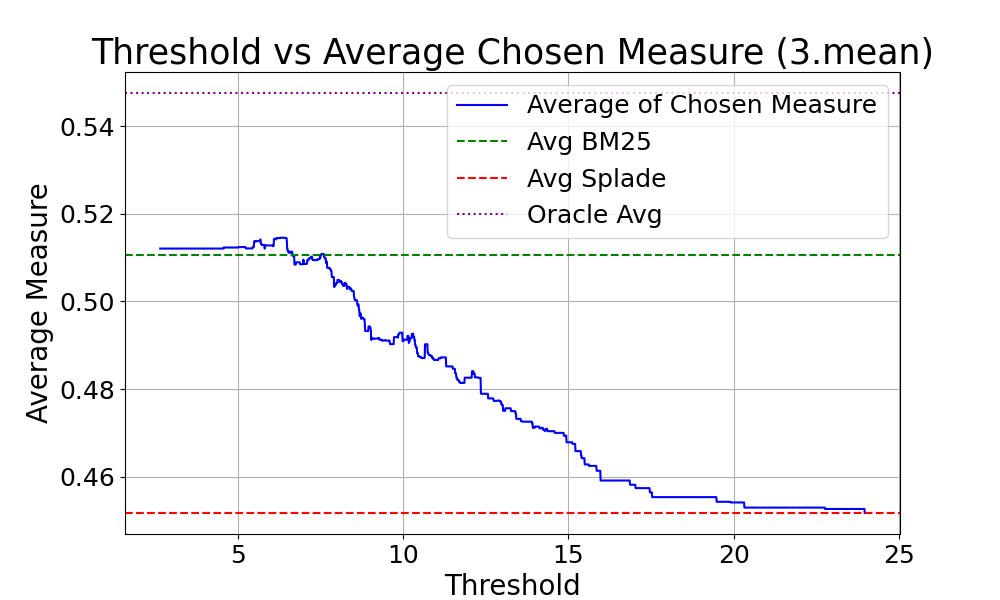}
        \caption{BM25 vs SPLADE - Based on Letor feature L.BM25}
        \label{fig:subfig3}
    \end{subfigure}
    \hspace{-1mm} 
    \begin{subfigure}[t]{0.49\columnwidth} 
        \centering
        \includegraphics[width=\textwidth]{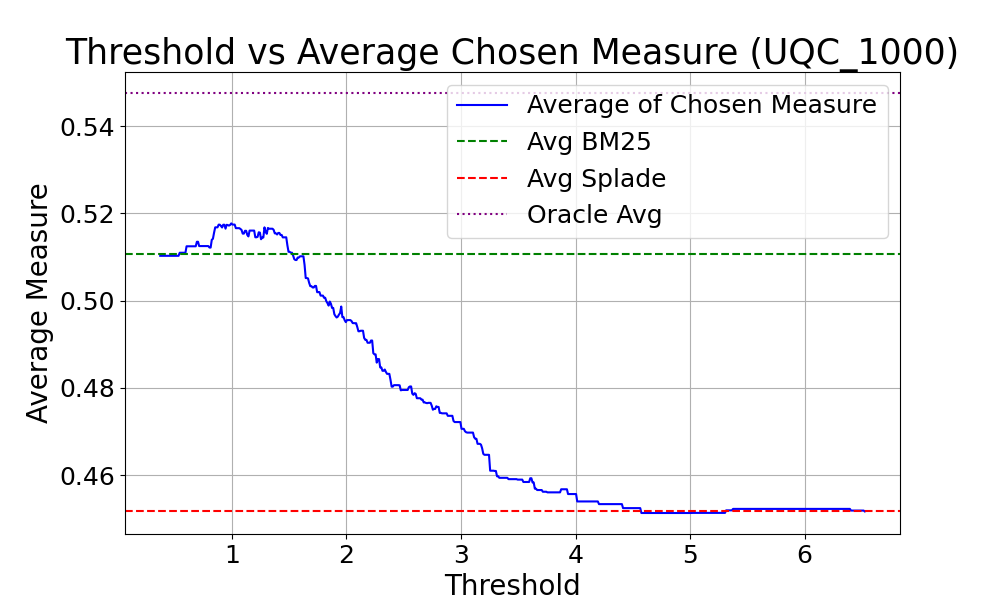} 
        \caption{BM25 vs SPLADE - Based on SOTA UQC} 
        \label{fig:subfig4}
    \end{subfigure}
    \caption{The automatic ranker selection based on a single QPP seldom outperform the individual system - ROBUST collection with NDCG. }
    \label{fig:SelectiveRankingModel}
\end{figure}

Figure~\ref{fig:SelectiveRankingModel} presents NDCG results for two scenarios: an automatic selective process between the BM25 and BM25 with QE rankers (top Sub-Figures) and between the BM25 and SPLADE rankers (bottom Sub-Figures). We evaluate the LETOR feature L.BM25 alongside the SOTA UQC predictor from BM25 ranking. While the threshold value could be learned within a train/test framework, the experiments illustrated in Figure~\ref{fig:SelectiveRankingModel}  explore only a predefined range of threshold values, spanning from the minimum to the maximum QPP value, with a step of 0.01. The results indicate  UQC is a more effective predictor than  L.BM25 for ranker selection while the correlation values were the other way around. For most threshold values the meta-system does not outperform the best stand-alone system. The observed increase in NDCG is modest, about 4\%.

We then consider an alternative scenario where the predicted performance values for both the ranker with QE and the ranker without QE are available. In this case, the decision to apply QE is made by directly comparing the predicted values of the two configurations. This approach assumes that the QPP can estimate the performance of both configurations with sufficient accuracy. 
This hypothesis leverages the relative strength of QPP predictions to dynamically select the optimal configuration, thereby maximizing retrieval effectiveness. Unlike the threshold-based strategy discussed earlier, this method streamlines the decision-making process by directly choosing the configuration with the higher predicted performance.

Let $P_{\text{R1}}(q)$ denote the predicted performance value for the R1 ranker for query $q$, $P_{\text{R2}}(q)$ the predicted performance value for the R2 ranker for query $q$, $R_{\text{R1}}(q)$ and $R_{\text{R2}}(q)$ the two alternative ranker, and $R_{\text{opt}}(q)$ the optimal ranker selected for $q$.
The optimal configuration is  as follows:
\begin{equation}
\label{eq:Selective2}
R_{\text{opt}}(q) =
\begin{cases} 
R_{\text{R2}}(q), & \text{if } P_{\text{R2}}(q) > P_{\text{R1}}(q), \\
R_{\text{R1}}(q), & \text{otherwise}.
\end{cases}
\end{equation}

In this framework, on the experimental part, the selective process failed, as the same configuration was chosen for all queries. This issue may stem from the normalization method applied during QPP calculation (see Section~\ref{subsec:feature}, Eq.~\ref{eq:Normalization}). Although alternative normalization approaches were tested, the problem persisted.

\begin{wraptable}{r}{0.5\textwidth} 
\centering
\caption{NDCG for the automatic selective query processing. Selection is between BM25 and BM25 QE (2nd col.) and between BM25 and SPLADE (3rd col.). The training uses the predictors (1st col.) and SVM -best- algorithm. $\blacktriangle$ indicates the selective process outperforms both individual rankers whose performance is reported in the top two rows.}
\begin{tabular}{l|c|c}
\hline
\textit{ROBUST} & BM25 vs & BM25 vs \\
\textit{NDCG} & BM25 QE & SPLADE \\
\hline
R1: BM25 & 0.5106 & 0.5106 \\
R2: BM25 or SPLADE  & 0.5322 & 0.4517 \\
Oracle selection & 0.5393 & 0.5476 \\
\hline
UCQ\footnotemark[1]  & 0.5325$\blacktriangle$ & 0.5108$\blacktriangle$ \\ 
L.BM25 & 0.5322  & 0.5108$\blacktriangle$  \\
LETOR & 0.5323$\blacktriangle$  & 0.5109$\blacktriangle$ \\
SOTA & 0.5322   & 0.5098 \\
All  & 0.5322 & 0.5121$\blacktriangle$ \\
\hline
\end{tabular}
\label{tab:SelectiveRankingModel}
\footnotetext[1]{We use the values calculated on BM25 run for all.}
\end{wraptable}

In a third attempt, we developed a model trained on past queries to make decisions for unseen queries. Using the $Q_A$ and $Q_{\overline{A}}$ framework outlined in Section~\ref{Subsec:training}, we experimented with both individual features and their combinations.


For  ROBUST, Table~\ref{tab:SelectiveRankingModel} presents results for the SVM model, which proved to be the most effective for selective query processing, outperforming each individual ranker. We evaluated BM25 in combination with either BM25 QE or SPLADE. BM25 alone achieved 0.5106, while BM25 QE reached 0.5322. An Oracle selecting the best ranker for each query attained 0.5393, slightly surpassing BM25 QE. A model trained on UQC achieved 0.5325. When combining BM25 and SPLADE, the Oracle achieved 0.5476, with individual ranker performances of 0.5106 for BM25 and 0.4517 for SPLADE. A trained model obtained 0.5121, slightly exceeding the performance of BM25 alone -3rd and 4th digit only, not statistically significant.

\section{Related Work}
\label{Sec:RelatedWork}
Early QPP methods can be categorized into pre-retrieval and post-retrieval approaches. Pre-retrieval predictors estimate query difficulty based on query characteristics and corpus statistics before any retrieval occurs~\cite{cronen2002predicting,Mothe2005Linguistics,hauff2008survey}. These methods are computationally efficient but often lack the granularity required for accurate performance estimation. Notable pre-retrieval predictors include query length, inverse document frequency (IDF), and term specificity metrics~\cite{he2004inferring,zobel1998reliable}. Extensions to pre-retrieval approaches have explored query variations to improve robustness~\cite{scells2018query} and the integration of external knowledge sources like Wikipedia~\cite{katz2014wikipedia}.

Post-retrieval predictors, on the other hand, analyse  retrieved documents to assess performance~\cite{carmel2010estimating,roitman2017enhanced,shtok2012predicting}, considering features such as the distribution of  retrieval scores or the coherence of the top-ranked documents. Notable methods include Normalized Query Commitment (NQC), Unnormalized Query Commitment (UQC), Weighted Information Gain (WIG), and Query Feedback (QF), which are widely used in recent QPP studies~\cite{faggioli2023query}. Ensemble-based methods have also been explored, combining multiple predictors to improve accuracy and robustness~\cite{grivolla2005automatic,zhou2007query,raiber2014query,chifu2018query,mizzaro2018query,ROY20191026}. 

With neural IR models, such as BERT-based retrievers, was introduced new challenges for traditional QPP methods. These models leverage contextual embeddings, which differ fundamentally from the term-frequency-based representations of sparse rankers like BM25. Studies have shown that traditional QPPs often underperform in dense retrieval settings~\cite{FaggioliEtAl2023b}. Recent research has proposed predictors tailored to dense rankers, such as coherence-based predictors that leverage embedding representations~\cite{vlachou2023coherencebasedpredictorsdensequery} or based on injecting noise into contextualized neural representations ~\cite{10.1145/3583780.3615270}. While these approaches demonstrate improved performance on datasets like TREC Deep Learning Track, their generalizability across retrieval paradigms remains an open question. Other advances in post-retrieval QPP include modelling retrieval coherence through document association networks~\cite{arabzadeh2021retrievalcoherency}, leveraging neural embedding representations for coherence-based predictors~\cite{vlachou2023densepredictors}, and exploring QPP applications in conversational search contexts~\cite{meng2023conversationalqpp}.

QPP has also been applied to various downstream tasks, including selective query processing, query expansion, and conversational search. For instance, effective QPP can guide systems in determining whether to apply query expansion or adjust configurations to improve retrieval effectiveness. 
In conversational contexts, QPP helps guide subsequent interactions, enhancing user experience~\cite{meng2023conversationalqpp,radlinski2017conversational}. Furthermore, QPP has shown potential in dynamic ranker selection, aiding retrieval systems in choosing the best ranker for a given query~\cite{mothe2023selective}. Additional applications include session search~\cite{ustinovskiy2013sessionbasedqpp} and the automatic detection of query intent~\cite{zamora2014queryintentdetection}. 

While previous studies have examined QPP in isolation for either sparse or dense rankers, few have conducted a comprehensive cross-paradigm analysis. Few also have considered multiple collections.  Additionally, the effectiveness of QPP-driven applications, such as selective ranker selection or query expansion, has received limited attention. 
This paper addresses these gaps. 

\section{Conclusion}
\label{Sec:Conclusion}
In this paper, we provide  a comprehensive cross-paradigm evaluation considering traditional QPP methods (NQC, UQC, WIG and QF) as well as aggregated LETOR features, and newer dense-based predictors (Bert$_{Bi}$, Bert$_{Cross}$)  across sparse (BM25, DFree), query-expansion-enhanced retrieval systems (BM25 QE, DFRee QE), and dense (SPLADE, ColBERT) rankers. To the best of our knowledge this is the first time dense ranker performance are reported in that way for earlier TREC collections. 

We highlight the limitations of existing QPP methods, particularly their inability to generalize across datasets and rankers, using robust experimental evidence. We include and analyze embedding-aware QPP features, demonstrating their potential for dense rankers but also identifying challenges in achieving consistent performance across collections.
Moreover, most of the reports on studies have no room to report plots in addition to correlation and p-values. For example on the teaser figure of this paper, the latest subfigure is a correlation of $0.288^\ddagger$, but we can see that the link between the two variables (MRR@10 and Bert$-Bi$) is very weak. 

Finally, our findings reveal that QPP-guided applications, such as selective query processing, offer limited gains in practice, emphasizing the need for more robust and adaptable QPP metrics.

These contributions provide new insights into the fragility of current QPP methods and lay the groundwork for developing more generalizable and effective predictors for IR.

On the lessons learnt from this study, we would recommand to assess QPPs on various collection types, to plot the correlations, even when found statistically significant, and eventually, depending on the objectives, to assess them with different  performance measures and rankers.

\newpage
\bibliographystyle{unsrt}  
\bibliography{main}

\end{document}